\documentclass[american,aps,prd,superscriptaddress,10pt,twocolumn,nofootinbib,preprintnumbers]{revtex4-2}

\usepackage[T1]{fontenc}
\usepackage[utf8]{inputenc}
\usepackage{amsmath,amssymb,bm}
\usepackage[dvipdfmx]{graphicx}
\usepackage{xcolor}
\usepackage{hyperref}



\usepackage{babel}
\begin{document}

\preprint{YITP-26-90, RIKEN-iTHEMS-Report-26}

\title{Evading Cosmological Strong Coupling in Non-minimally Coupled Vector Gravity}

\author{Antonio De Felice}
\affiliation{Center for Gravitational Physics and Quantum Information, Yukawa Institute for Theoretical
Physics, Kyoto University, 606-8502, Kyoto, Japan}

\author{Seishi Enomoto}
\affiliation{Faculty of Science and Engineering, Kyushu Sangyo University,
Fukuoka 813-8503, Japan}
\affiliation{Department of Physics, Faculty of Engineering Science, Yokohama National University, Yokohama 240-8501, Japan}

\author{Nagisa Hiroshima}
\affiliation{Department of Physics, Faculty of Engineering Science, Yokohama National University, Yokohama 240-8501, Japan}
\affiliation{RIKEN Center for Interdisciplinary Theoretical and Mathematical Sciences(iTHEMS), RIKEN, Wako 351-0198, Japan}

\author{Atsushi Naruko}
\affiliation{National Institute of Technology, Fukui College, Sabae, Fukui 916-8507, Japan}
\affiliation{Department of Physics, Faculty of Engineering Science, Yokohama National University, Yokohama 240-8501, Japan}
\affiliation{Asia Pacific Center for Theoretical Physics, Pohang 37673, Korea}
\affiliation{Center for Gravitational Physics and Quantum Information, Yukawa Institute for Theoretical
Physics, Kyoto University, 606-8502, Kyoto, Japan}

\date{\today}

\begin{abstract}
Recent analyses of Proca theories with non-minimal curvature couplings have
uncovered an additional scalar degree of freedom with an identically
vanishing propagation speed, $c_s^2=0$, signaling a scale-dependent
strong-coupling problem. In this {\it Letter}, we show that an additional
derivative interaction can lift this zero-speed degeneracy and restore a
non-degenerate quadratic dynamics for the scalar perturbations. In an open
region of parameter space, the scalar kinetic matrix is positive definite
and the high-frequency propagation speeds are real and positive. At
quadratic order, this removes the specific signature of strong coupling
associated with the $c_s^2=0$ mode. We also uncover a non-uniform limit as 
the temporal vector condensate approaches zero: at fixed nonzero $A_0$,
the formal extreme-ultraviolet regime develops a ghost, while the kinetic
structure of the exactly vanishing-condensate branch is different. The
momentum scale at which this ghost appears is pushed toward increasingly
high values as $A_0\to0$.
Whether the ghost scale ultimately lies above the EFT cutoff depends on an independent determination of the EFT cutoff.
Finally,
we identify a stable de Sitter fixed point with $A_0\neq0$, surrounded by
a finite region in which the scalar no-ghost and high-frequency stability
conditions remain satisfied.
\end{abstract}

\maketitle

\emph{Introduction.---}
The origin of late-time cosmic acceleration remains one of the central
open problems in modern physics. Although a cosmological constant provides
an economical phenomenological description, its fine-tuning problems have
motivated the search for dynamical alternatives, in particular modified
gravity and dark-energy models with additional degrees of freedom. Among
them, vector-tensor theories are especially appealing: vector fields carry
intrinsic derivative and directional structures, and have been used both in early-universe inflation~\cite{Ford:1989me} and as candidates for isotropic late-time acceleration~\cite{Armendariz-Picon:2004whg}, provided that anisotropies are dynamically controlled~\cite{Barrow:2005qv,Maleknejad:2012fw}.

The price of this richness is theoretical fragility. Non-minimal couplings
between a vector field and curvature, or between vector fields and matter
sectors, may easily activate ghost or gradient instabilities in cosmological
perturbations~\cite{BeltranJimenez:2013btb}. Recent works have explored scenarios
in which certain ghost degrees of freedom may be compatible with unitary
time evolution under special assumptions~\cite{Ewasiuk:2026ghosts,
Deffayet:2026unitary,Deffayet:2026counterexamples}. 
Here we take the more conservative route: a viable cosmology must contain no ghost in its physical spectrum. 
This requirement is naturally addressed in generalized Proca
theories~\cite{Heisenberg:2014rta}, which classify derivative
vector self-interactions propagating the correct number of
degrees of freedom and admit cosmologically viable,
perturbatively stable solutions~\cite{DeFelice:2016yws},
while screening fifth forces on local scales~\cite{DeFelice:2016cri}.

However, this expectation has recently been challenged. In non-minimally coupled
Proca cosmology, an additional scalar mode appears on a smooth Friedmann-Lema\^itre-Robertson-Walker (FLRW) 
background with identically vanishing propagation speed,
$c_s^2=0$ or disappears at the level of linear perturbation around the FLRW universe, signaling the strong coupling~\cite{DeFelice:2025ykh}. A closely related phenomenon occurs in non-minimal three-form cosmology, where the dynamics reduces to an emergent
cuscuton-like constraint~\cite{DeFelice:2025threeform}. 
Such zero-speed modes signal the onset of scale-dependent
strong coupling, implying that the cutoff of the effective
theory becomes momentum-dependent
\cite{Arkani-Hamed:2003juy,Cheung:2007st}, 
casting doubt on whether non-minimally coupled gauge fields can provide a
controlled cosmological effective field theory (EFT).  
Here we pursue a complementary possibility. Rather than enforcing the
degeneracy that removes the additional scalar mode, we ask whether this mode can become genuinely dynamical and healthy once the vector sector is enlarged by a divergence-squared interaction.

In this {\it Letter} we show that this obstruction is not fatal. We study a
four-parameter class of non-minimally coupled vector-tensor theories,
labelled by $(\xi_1,\xi_2,\xi_3,\xi_4)$, and show that derivative
interactions, in particular the $(\nabla^\mu X_\mu)^2$ operator governed by
$\xi_3$, can lift the zero-speed scalar mode. As a consequence, there exist regions of
parameter space where the homogeneous background is stable and the scalar
sector is free from ghosts and gradient instabilities, with real and
positive propagation speeds.

Another indication we obtain is that 
the limit in which the temporal condensate
$A_0$ approaches zero is non-uniform: the formal extreme-UV limit
at fixed nonzero $A_0$ does not commute with the exactly vanishing-condensate branch. 
The formal UV ghost is pushed to increasingly high momenta as
\(A_0\to0\). Whether it lies within the physical regime of the theory,
therefore depends on the cutoff scale of the effective theory 
whose explicit determination is beyond
the scope of the present analysis~\cite{Burgess:2003jk}.
Numerical integrations exhibit a finite basin of trajectories converging toward the healthy de Sitter fixed point.

\emph{Theory and background.---}
We study the action
\begin{align}
S&=\int {\rm d}^4x\sqrt{-g}\Big[\frac{M_{\rm Pl}^2}{2}(R-2\Lambda)-\frac14 F_{\mu\nu}F^{\mu\nu}
-\frac{\xi_1}{2}RX_\mu X^\mu
\nonumber\\
&-\frac{\xi_2}{2}R_{\mu\nu}X^\mu X^\nu
-\frac{\xi_3}{2}(\nabla_\mu X^\mu)^2
-\frac{\xi_4}{2}m_X^2X_\mu X^\mu\Big] ,
\label{eq:action}
\end{align}
where $F_{\mu\nu}=\partial_\mu X_\nu-\partial_\nu X_\mu$ and $\Lambda$ denotes the cosmological constant. 
Note that $\xi_3 (\nabla_\mu X^\mu)^2$ is the new term, which solves the problems found in ~\cite{DeFelice:2025ykh}.
We take a spatially flat FLRW line element with lapse $N(t)$ and scale factor $a(t)$, and a homogeneous vector profile
\begin{equation}
X_\mu{\rm d} x^\mu=-N(t)A_0(t){\rm d} t .
\end{equation}
After varying the action we set $N=1$ and define $H=\dot a/a$.
For a FLRW background, the variation of the action w.r.t. $A_0(t)$ yields 
\begin{align}
 &\xi_3 \ddot{A}_0 +3 H \xi_3 \dot{A}_0 
 + \Big[ 3 (2 \xi_1+\xi_2+\xi_3) \dot{H} \nonumber\\
 &\quad +3 (4 \xi_1+\xi_2) H^2 + \xi_4 m_X^2 \Big] A_0  = 0 .
 \label{eq-A}
\end{align}
We note that the equation for $A_0$ is linear in $A_0$ since the action is quadratic in $A_0$.
The Friedmann equation can be written as
\begin{align}
 &3 H^2 \left[ M_{\rm Pl}^2 + \left( \xi_1 + \xi_2 + \frac{3}{2} \xi_3 \right) A_0^2 \right] 
 + \frac{\xi_3}{2} \dot{A}_0^2 
 \notag\\
 & \quad
+ 3 (2 \xi_1+\xi_2+\xi_3) H \dot{A}_0 A_0
= M_{\rm Pl}^2\Lambda - \frac{\xi_4 m_X^2}{2} A_0^2 .
\label{eq-H}
\end{align}

A de Sitter solution with $\dot H=0$ and $\dot A_0=0$ simplifies the equation of motion for $A_0$ as follows
\begin{align}
&\left[ 3 (4\xi_1+\xi_2) H^2 +\xi_4m_X^2\right]A_0= 0 .
\label{eq:branch-condition}
\end{align}
This gives two candidate branches, 
\begin{equation}
\xi_4 \, m_X^2=-3H^2(4\xi_1+\xi_2).  
\label{eq:branch1}
\end{equation}
and $A_0=0$.

\emph{Scalar perturbations.---}
For the scalar sector we write
\begin{align}
{\rm d} s^2=&-N(t)^2(1+2\alpha){\rm d} t^2+2N(t)\partial_i\beta\,{\rm d} t{\rm d} x^i
\nonumber\\
&+a^2(t)(1+2\mathcal{R})\delta_{ij}{\rm d} x^i{\rm d} x^j+2\partial_i\partial_jE\,{\rm d} x^i{\rm d} x^j,
\end{align}
and
\begin{equation}
X_\mu{\rm d} x^\mu=-N(t)\,(A_0+\delta A_0)\,{\rm d} t+\partial_i\delta A\,{\rm d} x^i .
\end{equation}
We choose the flat gauge, $\mathcal{R}=E=0$, and redefine
\begin{equation}
\delta A_0=A_0\alpha+\delta\mathcal A_0 .
\end{equation}
The lapse function and the scalar shift perturbations yield nondynamical constraints. 
After moving to the Fourier space,
integrating them out gives a reduced quadratic action for
$\psi_i=(\delta\mathcal A_0,(k/a)\delta A)$ of the form
\begin{equation}
\mathcal L^{(2)}=K_{ij}\dot\psi_i\dot\psi_j
+L_{12}(\dot\psi_1\psi_2-\dot\psi_2\psi_1)-M_{ij}\psi_i\psi_j,
\label{eq:reduced-action}
\end{equation}
where $K_{ij}$ and $M_{ij}$ are symmetric matrices. 
We will not present the explicit form of the coefficients but $K_{ij}$ schematically looks like this:
\begin{equation}
 K_{ij} = a^3 \frac{c_1 (k/a)^2 A_0^2 + c_2}{c_3 (k/a)^2 A_0^2 + c_4},
\end{equation}
 where $c_{1,2,3,4}$ are functions of $\xi_i, H, \dot{A}_0$ and $A_0$ without any $k$-dependence. The crucial feature is that $k$ is always accompanied by $A_0$. 
 And hence, we will lose any $k$-dependence in $K_{ij}$ when $A_0$ exactly vanishes.
 
For a two-dimensional real and symmetric kinetic matrix \(K_{ij}\),
the absence of ghost modes requires \(K_{ij}\) to be positive definite.
By Sylvester's criterion, this is equivalent to demanding the positivity of
the leading principal minors, namely
\footnote{
For a \(2\times2\) real symmetric matrix, the conditions
\({\mathrm{Tr}}(K)>0\) and \(\det K>0\) are also equivalent to positive
definiteness. The form used in Eq.~\eqref{eq:noghost-general},
\(K_{11}>0\) and \(\det K>0\), is the standard Sylvester criterion in the
chosen basis.
}
\begin{equation}
K_{11}>0,\qquad \det K>0 .
\label{eq:noghost-general}
\end{equation}
In the high-momentum regime, for the branch with non-vanishing vector
condensate, \(A_0\neq0\), these two conditions reduce to
\begin{align}
K_{11}
&=
a^3\frac{(2\xi_1+\xi_2)(2\xi_1+\xi_2+2\xi_3)}
{2\xi_3},
\label{eq:K11}
\\
\det K
&=
-a^6\frac{H^2}{\xi_3A_0^4}
\left[\xi_3M_{\rm Pl}^2-\Xi A_0^2\right]
\left[M_{\rm Pl}^2 + (\xi_1+\xi_2)A_0^2\right],
\label{eq:detK}
\end{align}
where
\begin{equation}
\Xi
=
\frac{3}{2} (2\xi_1+\xi_2)^2 + (5\xi_1+2\xi_2)\xi_3 .
\label{eq:Xi}
\end{equation}
Equations~\eqref{eq:K11} and \eqref{eq:detK} therefore impose
constraints on the parameters of the theory and, in general, also on the
background value of the vector condensate. Since these conditions are obtained
in the high-\(k\) limit, their violation signals the presence of a UV ghost in the scalar sector.

It is instructive to consider a region of parameter space close to,
but not exactly on, the branch \(A_0=0\). Namely, let us take
\begin{equation}
0<\frac{|A_0|}{M_{\rm Pl}}\ll1 .
\end{equation}
This is a relevant situation because generic initial conditions need not
place the system exactly on the \(A_0=0\) branch. For any non-zero but small
\(A_0\), the high-momentum limit is controlled by the \(A_0\neq0\) expressions
above. In this regime one finds
\begin{equation}
\det K
\simeq
-a^6 H^2 \frac{M_{\rm Pl}^4}{A_0^4},
\qquad
0<\frac{|A_0|}{M_{\rm Pl}}\ll1 .
\end{equation}
Thus, for any finite non-zero condensate close to the $A_0=0$ branch, the determinant of the kinetic matrix becomes negative in the sufficiently high-momentum regime. Equivalently, there exists a momentum scale above which the UV asymptotics is governed by the ghostly sign of Eq.~\eqref{eq:detK}.

The scale at which this UV ghost becomes visible is pushed to larger and
larger momenta as \(A_0\to0\). Therefore, by tuning the initial value of
\(A_0\) sufficiently close to zero, one may push this scale above the cutoff of
the EFT, so that the ghost lies outside the regime of validity of the EFT. 
In this sense, the branch \(A_0=0\) is approached in a
singular way: the limits \(A_0\to0\) and \(k\to\infty\) do not commute.

Exactly at \(A_0=0\), however, the scalar kinetic matrix becomes diagonal,
with
\[
K_{11}=-\frac{\xi_3}{2} a^3,
\qquad
K_{22}=\frac{1}{2}a^3,
\]
and therefore
\[
\det K
=
-\frac{\xi_3}{4} a^6.
\]
The no-ghost condition then reduces simply to
\[
\xi_3<0 .
\]
This shows explicitly that the exactly vanishing-condensate branch is regular
and ghost-free for \(\xi_3<0\), even though an arbitrarily small but non-zero
condensate probes a different UV asymptotic regime.

\emph{Propagation speeds.---}
We perform a physical UV test by taking the WKB approximation. In the WKB regime, with $\psi_i=e^{-i\omega t}\tilde\psi_i$ and slowly varying coefficients, the reduced equations imply
\begin{equation}
    \omega^4+\frac{\mathcal{B}}{\mathcal{A}}\omega^2+\frac{\mathcal{C}}{\mathcal{A}}=0,
\label{eq:dispersion}
\end{equation}
where
\begin{align}
\mathcal{A}&=K_{11}K_{22}-K_{12}^2,\\
\mathcal{B}&=2K_{12}M_{12}-K_{11}M_{22}-K_{22}M_{11}-L_{12}^2,\\
\mathcal{C}&=M_{11}M_{22}-M_{12}^2 .
\end{align}
Imposing the no-ghost condition, the positivity of $\mathcal{A}$ can be assured. 
The two squared frequencies are real and positive, provided
\begin{equation}
-\frac{\mathcal{B}}{\mathcal{A}}>0,
\qquad
\frac{\mathcal{C}}{\mathcal{A}}>0,
\qquad
{\mathcal{B}}^2-4{\mathcal{A}}{\mathcal{C}}\ge0 .
\label{eq:speed-criteria}
\end{equation}
In the high-$k$ limit, these conditions determine the squared propagation speeds $c_{s\pm}^2$. 

A simple explicit point in the healthy region of the first branch  defined in Eq.~\eqref{eq:branch1} is
\begin{align}  
\xi_1&=\xi_3=1,
\quad
\xi_2=0,
\quad
\frac{\bar{A}_0}{M_{\rm Pl}}=\sqrt{0.2},
\quad
\frac{\bar{H}}{M_{\rm Pl}}=0.1 .
\label{eq:healthy-point}
\end{align}
At this point, the two kinetic eigenvalues are positive and the two scalar propagation speeds are
\begin{equation}
c_{s+}^2\simeq1.14,
\qquad
c_{s-}^2\simeq0.20 .
\label{eq:speeds-example}
\end{equation}
Therefore, the scalar sector is free from ghosts and high-frequency gradient instabilities. The value $c_{s+}^2>1$ should not be confused with a ghost or a tachyonic UV instability; the latter would require a wrong-sign kinetic term or a negative squared frequency in the dispersion relation at high-$k$. 
Whether superluminal characteristics imply an obstruction to a standard Lorentz-invariant UV completion is a distinct question~\cite{Adams:2006sv}, not the instability question addressed here.

\emph{Tensor and vector modes.---}
The remaining sectors are straightforward. Tensor perturbations possess
\begin{equation}
G_T \equiv 1+(\xi_1+\xi_2)\frac{A_0^2}{M_{\rm Pl}^2},
\quad
c_T^2=\frac{1+\xi_1 (A_0^2/M_{\rm Pl}^2)}{G_T}.
\end{equation}
Vector perturbations have the canonical kinetic coefficient
$K_V=a^3/2$, and therefore introduce no additional no-ghost condition.
Their squared sound speed is
\begin{equation}
c_V^2=
\frac{1+\left(\xi_1+\xi_2+\xi_2^2/2\right)(A_0^2/M_{\rm Pl}^2)}
{G_T}.
\end{equation}
Thus, the non-scalar sectors are healthy provided
$G_T>0$, $c_T^2>0$, and $c_V^2>0$.
Note that when $\xi_2$ vanishes, $c_T$ and $c_V$ coincide with the speed of sound.

\emph{Homogeneous stability.---}
We now show that the same scalar-healthy region contains a stable homogeneous background. On the branch \eqref{eq:branch1}, a de Sitter fixed point with constant $\bar H$ and $\bar A_0$ exists if
\begin{align}
\xi_4 \left( \frac{m_X}{M_{\rm Pl}} \right)^2&=-3\bar H^2(4\xi_1+\xi_2),
\label{eq:xi4-fixed}
\\
\frac{\Lambda}{M_{\rm Pl}^2} &=3\bar H^2 
\left(1+\frac{-2\xi_1+\xi_2+3\xi_3}{2}\beta_A^2 \right) .
\label{eq:Lambda-fixed}
\end{align}
where $\beta_A\equiv\bar A_0/M_{\rm Pl}$.
Linearizing the background equations around
$A_0=\bar A_0+\Delta A_0$ and $H=\bar H+\Delta H$, and using $\mathcal{N}=\ln(a/a_*)$, one obtains a closed equation for $\Delta A_0$,
\begin{align}
0&=\Delta A_0'' + 3 \Delta A_0'
\nonumber\\
&+3 \beta_A^2 \frac{3(4\xi_1+\xi_2)(2\xi_1-\xi_2-3\xi_3)}{\xi_3 
- \Xi \, \beta_A^2}\Delta A_0,
\label{eq:homogeneous-perturbation}
\end{align}
where the prime denotes ${\rm d}/{\rm d}\mathcal{N}$
 and \(\Xi\) is defined in Eq.~\eqref{eq:Xi}.

For the point \eqref{eq:healthy-point}, the perturbation equation becomes
\(
\Delta A_0''+3\Delta A_0'+2\Delta A_0=0
\), and therefore
\begin{equation}
\Delta A_0=c_1e^{-\mathcal{N}}+c_2e^{-2\mathcal{N}}.
\label{eq:fixed-point-decay}
\end{equation}
Both homogeneous modes decay. The associated parameters are
\begin{equation}
\frac{\xi_4 m_X^2}{M_{\rm Pl}^2}=-0.12,
\qquad
\frac{\Lambda}{M_{\rm Pl}^2}=0.033.
\label{eq:example-xi4-Lambda}
\end{equation}
This fixed point is therefore simultaneously attractive in the homogeneous
sector and healthy in the high-frequency scalar sector.

\emph{Numerical basin.---}
To verify that the fixed point is not an isolated artifact, we evolve the
background equations in terms of \(\mathcal{N}\).  It is convenient to introduce
\(\mathfrak{A}_0\equiv A_0/M_{\rm Pl}\) and to work in Planck units.  Combining
Eqs.~\eqref{eq-A} and \eqref{eq-H} gives the closed equation
\begin{align}
\mathfrak{A}_0''+\frac{d_3 (\mathfrak{A}_0')^3+d_2 (\mathfrak{A}_0')^2+d_1 \mathfrak{A}_0'+d_0}{2 (2 \Lambda - \mu_X \mathfrak{A}_0^2) (\xi_3 - \Xi \,\mathfrak{A}_0^2)} = 0,
 \label{eq:A0-numerical}
\end{align}
where \(\mu_X\equiv \xi_4m_X^2\), \(\Xi\) is defined in Eq.~\eqref{eq:Xi}, and
 \begin{align}
 d_3 &= \xi_3 \left[ \mu_X \mathfrak{A}_0^2 (4 \xi_1+2 \xi_2+\xi_3)
   -2 \Lambda (2 \xi_1+\xi_2) \right], \\
d_2 &= 3 \mu_X \mathfrak{A}_0^3 \left[\xi_3 (8 \xi_1+5 \xi_2 +3 \xi_3)
+3 (2 \xi_1+\xi_2)^2 \right] \notag\\
& + 2 \mathfrak{A}_0 \left\{ \mu_X \xi_3 
- \Lambda \left[ \xi_3 (\xi_2-2 \xi_1)+3 (2 \xi_1+\xi_2)^2 \right] \right\}, \\
d_1 &= 3 \Big\{ \mu_X \mathfrak{A}_0^4 \left[ 6 \xi_2 (2 \xi_1+\xi_2)
 + \xi_3 (22 \xi_1+16 \xi_2 +9 \xi_3) \right] \notag\\
   & \qquad + 2 \Lambda \mathfrak{A}_0^2 \left[ 6 \xi_1 (2 \xi_1+\xi_2)
   + \xi_3 (2 \xi_1-\xi_2) \right] \notag\\
   & \qquad + 2 \mu_X \mathfrak{A}_0^2 (6 \xi_1+3 \xi_2+2 \xi_3) 
   + 4 \Lambda \xi_3 \Big\}, \\
d_0 &= 3 \mathfrak{A}_0 \left[ \mathfrak{A}_0^2 (2 \xi_1+2\xi_2+3 \xi_3)+2 \right] \notag\\
 & \times
\left\{ \mu_X \mathfrak{A}_0^2 (-2 \xi_1+\xi_2+3 \xi_3) 
+ 2 \mu_X + 2 \Lambda (4 \xi_1+\xi_2)  \right\}.
\end{align}
The corresponding Friedmann equation is
\begin{align}
H^2 =
\frac{2\Lambda-\mu_X \mathfrak{A}_0^2}
{\xi_3(\mathfrak{A}_0')^2+e_1\mathfrak{A}_0'\mathfrak{A}_0+e_2\mathfrak{A}_0^2+6}
\,,
\label{eq:H2-numerical}
\end{align}
with
\begin{align}
e_1=6(2\xi_1+\xi_2+\xi_3),
\quad
e_2=3(2\xi_1+2\xi_2+3\xi_3).
\label{eq:e1e2}
\end{align}
For the example in Eq.~\eqref{eq:healthy-point}, namely
\(\xi_1=\xi_3=1\), \(\xi_2=0\),
\(\bar A_0/M_{\rm Pl}=\sqrt{0.2}\), and \(\bar H/M_{\rm Pl}=0.1\),
Eqs.~\eqref{eq:example-xi4-Lambda} imply
\(\mu_X/M_{\rm Pl}^2=-0.12\) and \(\Lambda/M_{\rm Pl}^2=0.033\).
Trajectories in a finite neighborhood of this de Sitter point approach the
attractor while the scalar no-ghost quantities and the two high-frequency
squared speeds remain positive. This is the required open stability region:
the solution is neither protected by a singular tuning nor supported by a
vanishing kinetic eigenvalue. The denominator of Eq.~\eqref{eq:A0-numerical}
identifies the boundaries of this dynamical chart. For the numerical example
used here, \(2\Lambda-\mu_Xa_0^2\) has no positive zero, while
\(\xi_3-\Xi \, \mathfrak{A}_0^2=0\) gives \(a_0^2=1/11\), which lies outside the displayed
basin around \(\mathfrak{A}_0^2=0.2\).

\begin{figure}[t]
\centering
\IfFileExists{A0_H_to_fixed_point.pdf}{\includegraphics[width=0.88\columnwidth]{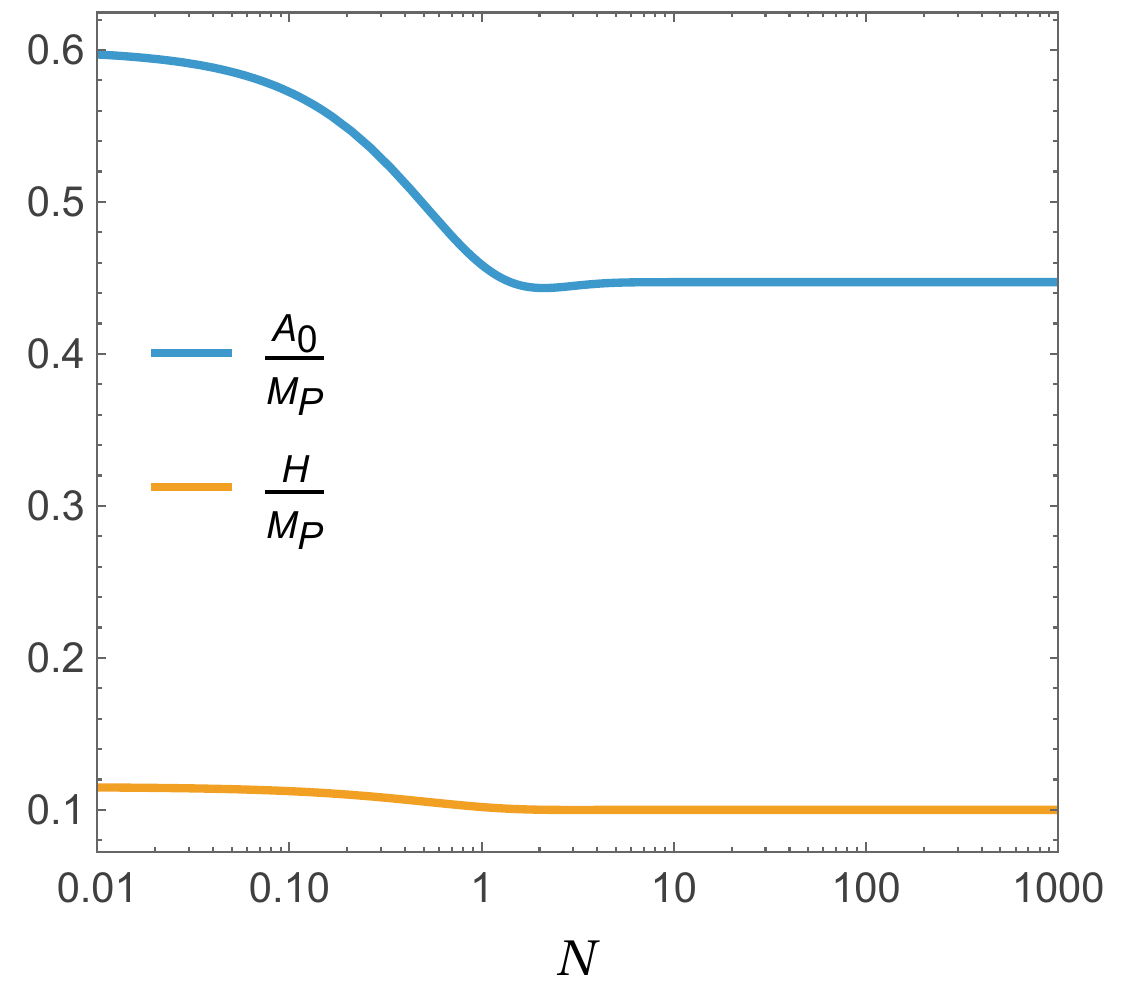}}{%
\IfFileExists{A0_H_to_fixed_point.png}{\includegraphics[width=0.88\columnwidth]{A0_H_to_fixed_point.png}}{\fbox{\parbox[c][3.2cm][c]{0.84\columnwidth}{\centering Background evolution}}}}

\vspace{4mm}
\IfFileExists{stability_to_fixed_point.pdf}{\includegraphics[width=0.88\columnwidth]{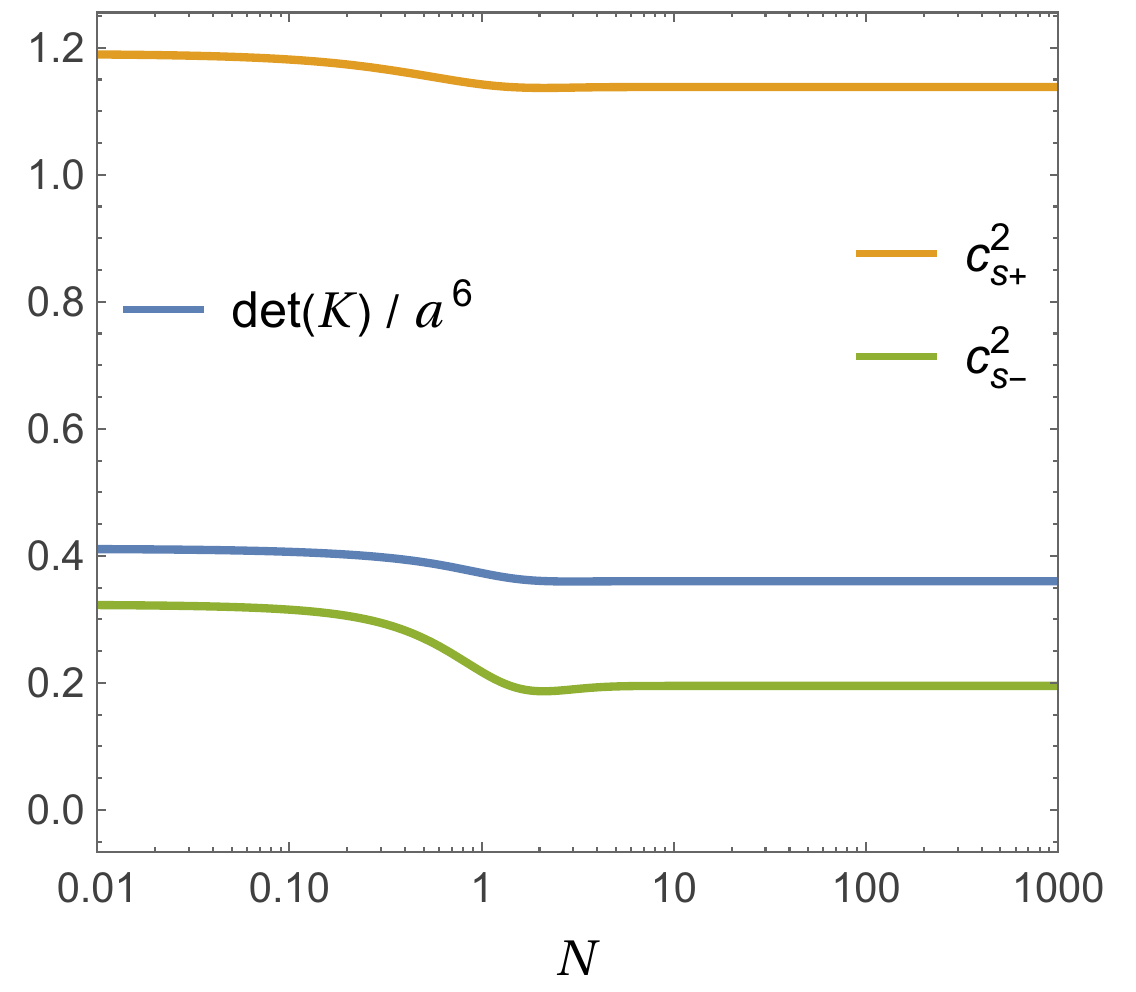}}{%
\IfFileExists{stability_to_fixed_point.png}{\includegraphics[width=0.88\columnwidth]{stability_to_fixed_point.png}}{\fbox{\parbox[c][3.2cm][c]{0.84\columnwidth}{\centering Stability diagnostics}}}}
\caption{Numerical basin for the fixed point of Eq.~\eqref{eq:healthy-point}.
Top: trajectories approaching the de Sitter point
$A_0/M_{\rm Pl}=\sqrt{0.2}$, $H/M_{\rm Pl}=0.1$. Bottom: high-frequency stability
diagnostics, including $(\det K)/a^6$ and $c_{s\pm}^2$. The remaining scalar
no-ghost condition is $K_{11}/a^3=4>0$. The vector kinetic coefficient is
$K_V/a^3=1/2$, while the tensor no-ghost condition gives
$G_T=1+(A_0/M_{\rm Pl})^2>0$. Since $\xi_2=0$ in this example, the tensor and
vector propagation speeds are luminal.}
\label{fig:basin}
\end{figure}

\emph{Conclusions.---}
We have shown that non-minimal vector-curvature couplings need not force the
additional scalar sector into a pathological UV regime. The key point
is that the divergence-squared interaction controlled by \(\xi_3\) changes the
scalar kinetic structure: in an open region of parameter space the kinetic
matrix is positive definite and the two high-frequency scalar propagation
speeds are real and positive.

The limit toward the exactly vanishing-condensate branch remains subtle. At
strictly \(A_0=0\), the scalar kinetic matrix is diagonal and ghost-free for
\(\xi_3<0\). For any finite but arbitrarily small \(A_0\), however, the formal
high-momentum limit is governed by a different asymptotic regime, and the two
limits \(A_0\to0\) and \(k\to\infty\) do not commute. This does not by itself
establish a physical instability within the effective theory, but it shows that
the cutoff scale around the condensate background is an essential part of the
complete viability question.

On the non-vanishing-condensate branch, we identified an open, non-fine-tuned region of parameter space in which all propagating modes are simultaneously free from ghosts and ultraviolet instabilities. The de Sitter fixed point in Eq.~\eqref{eq:healthy-point} provides an explicit cosmological realization within this healthy branch: the scalar no-ghost and high-frequency stability conditions, the tensor and vector stability conditions, and the homogeneous attractor condition are all satisfied in a finite neighborhood of phase space. This solution should therefore be regarded as a representative example, rather than as an isolated or uniquely viable configuration. It demonstrates explicitly that the extra scalar mode can remain genuinely dynamical and healthy without requiring singular tuning or a vanishing kinetic eigenvalue.

The broader lesson is conceptual. Extra gravitational degrees of freedom are
not automatically fatal; what matters is the simultaneous positivity of the
kinetic matrix, the UV dispersion relations, and the background
attractor structure, all evaluated on the same solution and in the same
parameter region. A natural next step is to determine the EFT cutoff around the
vector-condensate background and to confront the healthy region identified here
with the phenomenological constraints already studied in generalized Proca and
non-minimally coupled vector cosmologies~\cite{DeFelice:2016yws,DeFelice:2016uil,DeFelice:2017paw,DeFelice:2025ykh,DeFelice:2025threeform}.

\begin{acknowledgments}
The work of A.N. was partly supported by the JSPS KAKENHI Grant Numbers JP20H05852, JP23H01171, JP23K25868 and 26H02044. The work by N.H.  was supported in part by the JSPS KAKENHI Grant Numbers JP22K14035. 
The work of N.H, S.E, and A.N. is also supported in part by the MEXT Leading Initiative for Excellent Young Researchers Grant Number 2023L0013.

\end{acknowledgments}


%

\section*{Supplemental Material} 
\subsection*{Expansion of the action}

In the flat gauge,
\begin{align}
 \mathcal{R} = 0 \qquad \& \qquad E = 0 \,,
\end{align}
the scalar metric perturbations are given by
\begin{align}
 {\rm d} s^2
 &= -N(t)^2 (1+2\alpha){\rm d} t^2
 +2N(t)\partial_i\beta\,{\rm d} t\,{\rm d} x^i
 \notag\\
 &\qquad
 +a^2(t)\delta_{ij}{\rm d} x^i{\rm d} x^j\,,
\end{align}
while the matter perturbation is
\begin{align}
 X_\mu^{(1)}{\rm d} x^\mu
 =
 -N(t)\delta A_0{\rm d} t
 +\partial_i\delta A\,{\rm d} x^i\,.
\end{align}

Expanding the action up to quadratic order in perturbations, we obtain
\begin{align}
\mathcal{L}^{(2)}
=
\mathcal{L}^{(2)}
\bigl[
\alpha,\,
\beta,\,
\delta A_0,\,
\delta A
\bigr]
=
\mathcal{L}^{(2)}
\bigl[
\alpha,\,
\beta,\,
\delta\mathcal{A}_0,\,
\delta A
\bigr],
\end{align}
where
\begin{align}
\delta\mathcal{A}_0
=
\delta A_0
-
A_0\alpha.
\end{align}
By introducing $\delta\mathcal{A}_0$, the kinetic term for $\alpha$ disappears explicitly.
After integrating by parts, the time derivatives of $\alpha$ and $\beta$ disappear, so that they can be integrated out algebraically.

The Hamiltonian constraint equation schematically reads
\begin{align}
 h_1 \dot{\delta\mathcal{A}}_0
 + h_2 \delta\mathcal{A}_0
 + h_3 \dot{\delta A}
 + h_4 \delta A
 + h_5 \alpha
 + h_6 \beta
 =0\,,
\end{align}
while the momentum constraint equation is
\begin{align}
 m_1 \dot{\delta\mathcal{A}}_0
 + m_2 \delta\mathcal{A}_0
 + m_3 \dot{\delta A}
 + m_4 \delta A
 + m_5 \alpha
 + m_6 \beta
 =0\,.
\end{align}

Solving these constraint equations, one can express $\alpha$ and $\beta$ in terms of $\delta\mathcal{A}_0$ and $\delta A$.
Substituting these solutions back into the action, the quadratic action becomes a function only of $\delta\mathcal{A}_0$ and $\delta A$, as given in (\ref{eq:reduced-action}).

\subsection*{Derivation of (\ref{eq:dispersion})}

From the quadratic action (\ref{eq:reduced-action}), one obtains the equations of motion for $\delta\mathcal{A}_0$ and $\delta A$:
\begin{align}
\frac{{\rm d}}{{\rm d} t}
\left(
K_{11}\dot{\delta\mathcal{A}}_0
+
K_{12}\dot{\delta A}
+
L_{12}\delta A
\right)
+
M_{11}\delta\mathcal{A}_0
+
M_{12}\delta A
&=0\,,
\\
\frac{{\rm d}}{{\rm d} t}
\left(
K_{12}\dot{\delta\mathcal{A}}_0
+
K_{22}\dot{\delta A}
-
L_{12}\delta\mathcal{A}_0
\right)
+
M_{12}\delta\mathcal{A}_0
+
M_{22}\delta A
&=0\,.
\end{align}

Invoking the WKB approximation, these equations become
\begin{align}
(\omega^2K_{11}-M_{11})\delta\mathcal{A}_0
+
(\omega^2K_{12}
+i\omega L_{12}
-M_{12})
\delta A
&=0\,,
\\
(\omega^2K_{12}
-i\omega L_{12}
-M_{12})
\delta\mathcal{A}_0
+
(\omega^2K_{22}
-M_{22})
\delta A
&=0\,.
\end{align}

A non-trivial solution exists only if $\omega$ satisfies
\begin{align}
0
&=
(\omega^2K_{11}-M_{11})
(\omega^2K_{22}-M_{22})
\notag\\
&\quad
-
(\omega^2K_{12}
+i\omega L_{12}
-M_{12})
(\omega^2K_{12}
-i\omega L_{12}
-M_{12}),
\end{align}
which can be simplified to (\ref{eq:dispersion}).

\end{document}